\def\pks {PKS\,1246{\tt -}410}

\def\HI {H\kern0.1em{\sc i}} 
\def\radm {rad m$^{-2}$} 

\documentclass[useAMS,usenatbib]{mn2e}
\usepackage{psfig}

\begin{document}
\title{Fields and Filaments in the Core of the Centaurus Cluster}
\author[]
{\parbox[]{6.in} {G. B. Taylor$^1$, A. C. Fabian$^2$, G. Gentile$^1$, S. W. Allen$^3$,
C. Crawford$^2$ and J. S. Sanders$^2$\\
\footnotesize
1. University of New Mexico, Dept. of Physics and
Astronomy, Albuquerque, NM 87131, USA; gbtaylor@unm.edu\\ 
2. Institute of Astronomy, Madingley Road,
Cambridge CB3 0HA, acf@ast.cam.ac.uk; csc@ast.cam.ac.uk\\
3. Kavli Institute of Particle Astrophysics and Cosmology,
Stanford University, Stanford, CA 94305, USA; swa@stanford.edu\\
}}

\maketitle
\begin{abstract}

We present high resolution images of the Faraday Rotation Measure (RM)
structure of the radio galaxy \pks\, at the center of the Centaurus
cluster.  Comparison with H$\alpha$-line and soft X-ray emission
reveals a correspondence between the line-emitting gas, the soft X-ray
emitting gas, regions with an excess in the RM images, and signs
of depolarization.  Magnetic field strengths of 25 $\mu$G, organized
on scales of $\sim$1 kpc, and intermixed with gas at a temperature of
5 $\times 10^6$ K with a density of $\sim$0.1 cm$^{-3}$ can reproduce
the observed RM excess, the depolarization, and the observed X-ray
surface brightness.  This hot gas may be in pressure equilibrium with
the optical line-emitting gas, but the magnetic field strength of 25
$\mu$G associated with the hot gas provides only 10\% of the thermal
pressure and is therefore insufficient to account for the stability of
the line-emitting filaments.

\end{abstract}


\section{Introduction}

The Centaurus cluster, Abell 3526, is a nearby (redshift z=0.0104),
X-ray bright galaxy cluster.  At the center of this cluster is the
bright elliptical galaxy NGC 4696, hosting the moderately powerful radio
source, \pks.  We have been engaged in detailed studies of the X-ray
and radio emission from this cluster (Sanders \& Fabian 2002, Taylor,
Fabian \& Allen 2002), and have
recently obtained a further 200 ksec of Chandra data (Fabian et al. 2005).
The new Chandra image reveals a complex structure
within the central few kiloparsecs.  A plume-like structure swirls
clockwise to the NE and wraps around the radio source.  There are
clear signs of interaction between the X-ray and radio emission
including: (1) strong X-ray emission that matches the shape of the
radio emission just south of the core; (2) a faint rim of hard X-ray
emission along the northern edge of the radio source; and (3) deep
cavities in the X-ray emission on both the east and west sides.  We
also find a compact X-ray component coincident with the radio and
optical core (Taylor et al. 2006).

\begin{figure*}
\centerline{\psfig{figure=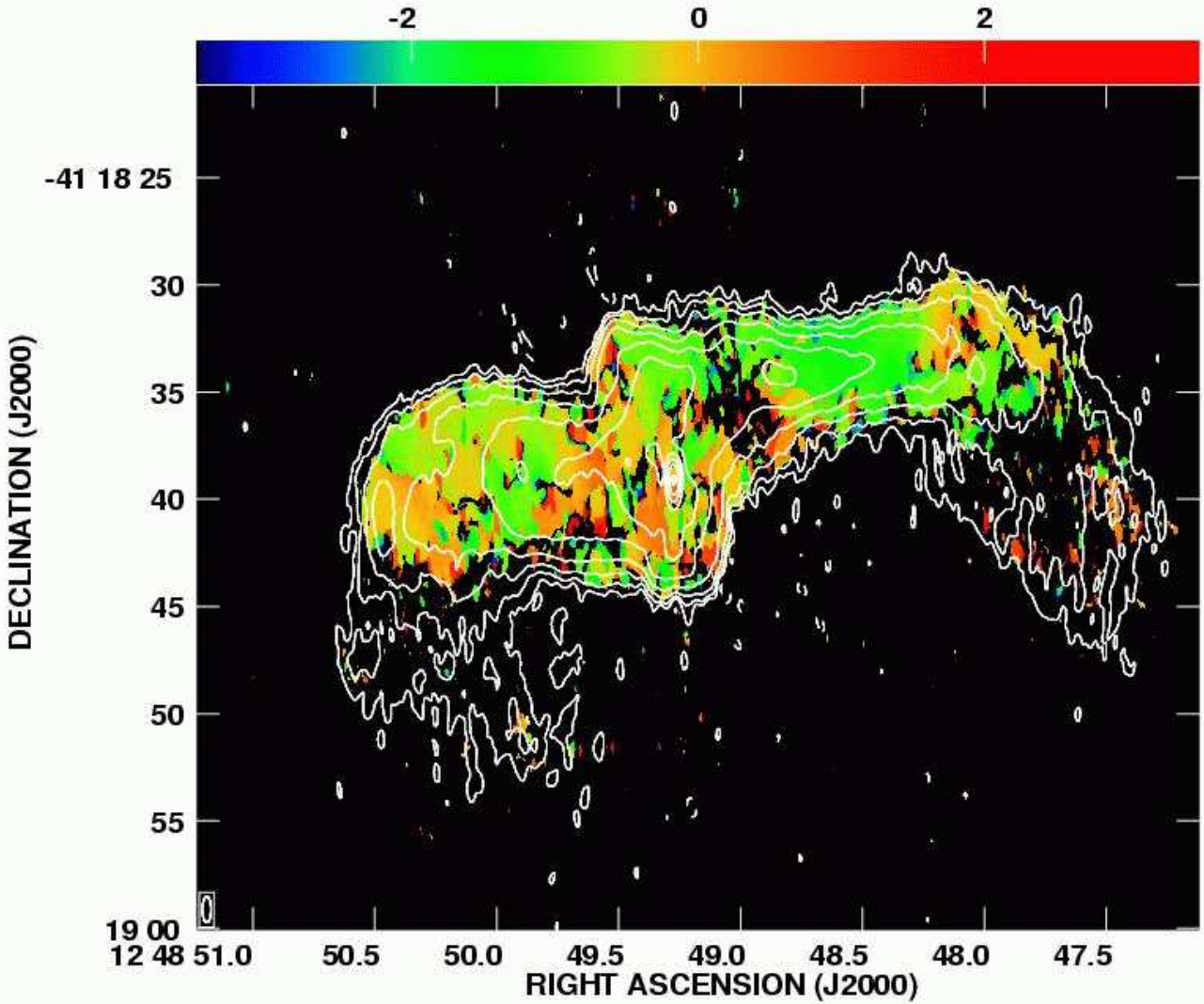,width=1.05\textwidth} }
\caption{The rotation measure (RM) map for PKS\,1246{\tt -}410 
with contours from the 5 GHz total intensity image overlaid. 
Contours start at 0.15 mJy/beam and increase by factors of 2.
The angular resolution is $1.2 \times 0.4$ arcsec.
\label{fig1}}
\end{figure*}

\begin{figure*}
\centerline{\psfig{figure=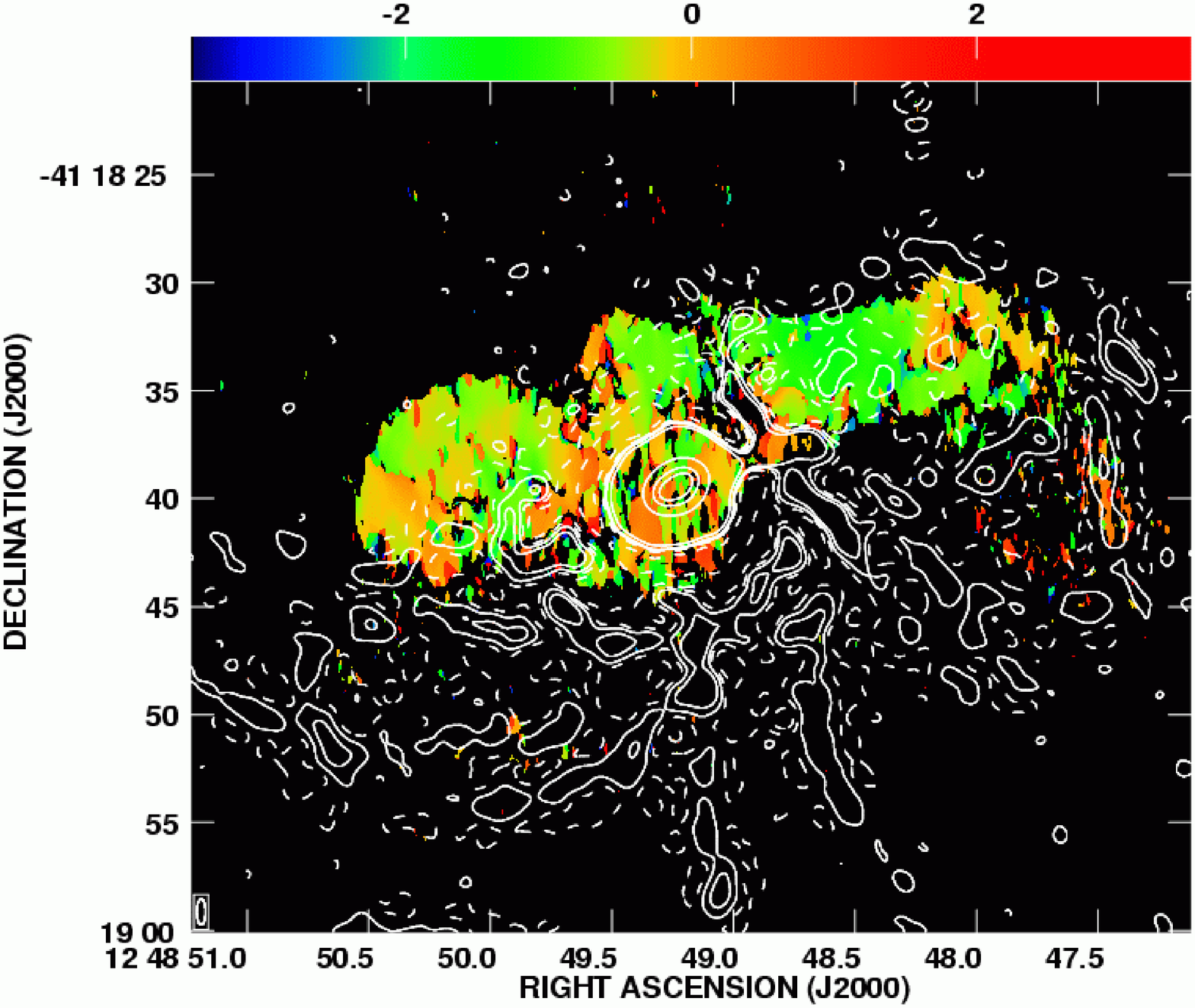,width=1.05\textwidth} }
\caption{The rotation measure (RM) map for PKS\,1246{\tt -}410 
with contours from the unsharp-masked H$\alpha$ image 
overlaid. The color bar ranges from $-$3500 to 3500 \radm.
Relative contours are drawn starting at 2.5$\sigma$ and 
increasing by factors of 4 (Crawford et al. 2005).  
\label{fig2}}
\end{figure*}

From VLA data primarily in BnA configuration, Taylor et al. (2002)
imaged the rotation
measure (RM) distribution across \pks\ at a resolution of 2.1 $\times$
1.2 arcsecond.  The RMs appear patchy and are
typically in the range $-$500 to +500 \radm, with a maximum absolute value of
$\sim$1800 \radm.  Assuming that the RMs are produced by cluster
magnetic fields as found for other galaxies at the center of dense
clusters (see review by Carilli \& Taylor 2002), simple magnetic field
topologies lead to estimates of the magnetic field strength of $\sim$11
$\mu$G.  The Centaurus cluster is exceptional in that the radio source
\pks\ provides a well resolved probe of the central 10 kpc of a
cooling core cluster.  

Taylor et al. (2002) suggested that some regions of higher RM lay
behind regions of high X-ray surface brightness.  This raised the
possibility that (1) some of the bright inner X-ray emission is 
located in front of the radio source; and (2) this thermal gas within
a few kiloparsecs of the radio could contribute significantly to the RM.
Rudnick \& Blundell (2003) used the observations of Taylor 
et al. to investigate the possibility of a local cocoon producing the
high RMs, and claimed to find evidence for this in a correlation
between the intrinsic polarization angles and the RM.  However, 
Ensslin et al. (2003) showed that the relationship was an artifact of
the analysis used, and a better formed statistic showed no correlation.
Here, we present improved observations of the RM structure in
\pks\ and further explore the evidence for local interactions through
detailed comparison with X-ray and H$\alpha$ images.

The arc-shaped structures seen in the H$\alpha$ filaments appear
quite smooth with no indication on subkiloparsec scales of
surrounding turbulence (Crawford et al. 2005).  This smoothness,
and the commonality of structure between optical filaments, X-ray
arcs, and the dust lanes suggest that these could all have been
shaped by strong magnetic fields.  An alternative explanation
could be a bulk laminar flow within the intracluster medium.
Detailed comparisons between the Faraday rotation measure and
the optical filaments should allow us to measure the strength
of the magnetic fields in the region of the filaments and 
thereby discriminate between these two possibilities. 

Here we present VLA (Very Large Array)\footnote{The National Radio
Astronomy Observatory is a facility of the National Science Foundation
operated under a cooperative agreement by Associated Universities,
Inc.} observations of the radio galaxy PKS\,1246{\tt -}410 at the
centre of the Centaurus cluster.  
In \S 2 we describe the radio observations.  In \S 3 we
summarize the H$\alpha$ emission, and in \S 4 we summarize
the X-ray observations.  In \S 5 we discuss the radio properties
of \pks\ and in \S 6 we compare the multi-wavelength
observations and consider the origin of the observed RMs.

We assume $H_0$ = 71 km s$^{-1}$ Mpc$^{-1}$ and $\Omega$
 = 1 so that 1\arcsec = 210 pc at the redshift (0.0104) of
the Centaurus cluster.
   
\section{Radio Observations and Data Reduction}

Observations of PKS\,1246{\tt -}410 were made at 4.635, 
4.885, 8.115, and 8.485 \,GHz at the VLA in its A configuration
on 2006 February 8.   The
source 3C\,286 was used as the primary flux density calibrator, and as
an absolute reference for the electric vector polarization angle
(EVPA).  Phase calibration was derived from the nearby compact source
J1316{\tt -}3338 with a cycle time between calibrators of about 16 minutes.
Instrumental calibration of the polarization leakage terms was
obtained using OQ\,208 which was observed over a wide range in
parallactic angle.  The data were reduced in AIPS (Astronomical Image
Processing System), following the standard procedures.  The AIPS task
IMAGR was used with a suitable taper to make Stokes $I$, $Q$ and $U$ images at each
of the 4 frequencies observed at the same resolution.  Polarized
intensity, $P$, images and polarization angle, $\chi$, images were
derived from the $Q$ and $U$ images.  The Faraday Rotation Measure (RM)
image was made from the combination of the $\chi$ images.
Pixels in the RM image were flagged when the error in any of the 
$\chi$ images exceeded 20 degrees.

\section{H$\alpha$ Observations}

A detailed study of the extended H$\alpha$ filaments surrounding
\pks\ has recently been published by Crawford et al. (2005).  They find 
a network of line-emitting filaments extending over the central 50
arcsec (10.5 kpc).  These filaments correlate with
arc-like plumes seen in soft X-ray emission.  Crawford
et al. speculate that both the H$\alpha$ filaments and the 
plumes of soft X-ray emitting gas have been drawn out of the
center by buoyant bubbles created by the central radio 
galaxy, \pks.

\section{X-ray Observations}

Fabian et al. (2005) presented deep (199 ksec) observations
of the Centaurus cluster, that showed well defined cavities
coincident with \pks.  Further evidence of outbursts from
the radio galaxy were seen in semi-circular edges to the 
east and west marked by sharp temperature increases.  

The electron density derived at the inner 70 kpc (4.2\arcmin) 
radius region, determined from a deprojection analysis, is
shown in Taylor et al. (2002).  The best-fitting $\beta-$model for the
density has parameter values $n_0 = 0.099\pm0.001$ cm$^{-3}$, $\beta =
0.39\pm0.01$, and a core radius, $r_c = 5.4\pm0.3$ kpc.  Sanders 
\& Fabian (2006) find a central temperature of $\sim$2 $\times 10^7$
K in the lobes.
Note that
the core size is similar to that of the radio source, which has made
`holes' in the X-ray emission at radii of 1--5~kpc.

\section{Radio Properties of \pks}
  
The radio source PKS\,1246{\tt -}410 is associated with NGC~4696, an
elliptical galaxy at the center of the Centaurus cluster (also known
as Abell 3526).  The radio power of PKS\,1246{\tt -}410 is 1.5 $\times
10^{24}$ W Hz$^{-1}$ at 1.6 GHz, identifying it as a low power
radio galaxy for which we would expect an FR~I morphology (edge dimmed
extended radio tails).  In fact the radio morphology at 5 GHz,
shown in Fig.~1, appears quite different from most low power radio 
galaxies (Parma et al. 1987).
The usual well defined core and thin jets feeding into larger lobes
are missing.  Instead there is a bright point source which we identify
as the core based on its central location and slightly flatter
spectral index, $\alpha = -0.5$ where $S_\nu \propto \nu^\alpha$
(Taylor et al. 2002).  The properties of the core were further
examined by Taylor et al. (2006) who found it to be radiatively
underluminous but powering a jet (detected by the VLBA) that 
carries off enough energy to inflate the cavities seen in the 
thermal gas.

The bright radio lobes of PKS\,1246{\tt -}410 are moderately well
polarized (4 -- 40\%), so it is possible to derive the RM distribution
across nearly the entire radio source.  This was done by Taylor et
al. (2002) at a resolution of 2.1 $\times$ 1.2 arcseconds.  The RMs
appear patchy and are not correlated with any features in total
intensity.  

We combine our new VLA-A configuration observations at 5 and 8.4 GHz
with the previous observations.  
Our higher angular resolution (1.2 $\times$ 0.4 arcsec) images reveal some fine-scale structure
in the Faraday rotation measure image (Fig.~1).  In general the
western lobe appears to have smoother RMs with typical values of 
$-$1000 \radm\ over a 5\arcsec\ region.  Much of the eastern lobe
is patchy with a scale size of $\sim$2\arcsec (0.4 kpc).  Typical RM gradients
are $\sim$400 \radm/arcsec.  Exceptionally large RM gradients are found
in the regions spatially coincident with H$\alpha$ filaments.  These 
regions are up to 2 arcseconds (0.4 kpc) wide and 5 arcsec (1 kpc) long.

Very little polarization is detected after the radio lobes
bend toward the south, but this is most likely due to a lack of
sensitivity since the lobes fade at 8.4 GHz to 100-300 $\mu$Jy/beam
and below.  Given our noise level of $\sim$25 $\mu$Jy we would only
detect polarization (at the 2 $\sigma$ level) if it was as high as
20-50\%.  There is a hint of an anti-correlation with line emission
across the southwest lobe, but more sensitive observations will be
required to confirm low fractional polarization there.

\section{Contributors to the Faraday Rotation}

We can rule out uniform internal Faraday rotation since it deviates
from a wavelength-squared law for changes in angle of more 
than 90 degrees (Burn 1966).  Internal Faraday rotation also predicts
significant depolarization of the radio
source since emission produced at the back interferes with emission
produced at the front.   The RM of 2000 \radm\ seen in \pks\ corresponds
to a change in angle of 400 degrees at 5 GHz.  

Faraday rotation from an external screen is thus likely and could have
contributions from (1) optical line-emitting filaments crossing in
front of the radio lobes; (2) the filaments of dense soft X-ray
emitting gas on 10-kpc scales; (3) a mixing layer between source
fields and the surrounding medium (Rudnick \& Blundell 2003); and 
(4) the 100-kpc scale cluster gas (Taylor et al. 2002).
The latter two explanations do not readily explain fine scale
RM structures correlated with the H$\alpha$ filaments so we
do not consider them further here.

For external Faraday rotation, the RMs are related to the density, $n_{\rm
  e}$, and magnetic field along the line-of-sight, $B_{\|}$, through
the cluster (or line-emitting gas) according to
$$ RM = 812\int\limits_0^L n_{\rm e} B_{\|} {\rm d}l ~{\rm
  radians~m}^{-2}~,
\eqno(1)
$$
where $B_{\|}$ is measured in $\mu$Gauss,  $n_{\rm e}$ 
in cm$^{-3}$ ~and d$l$ in kpc.  

\begin{figure}
\centerline{\psfig{figure=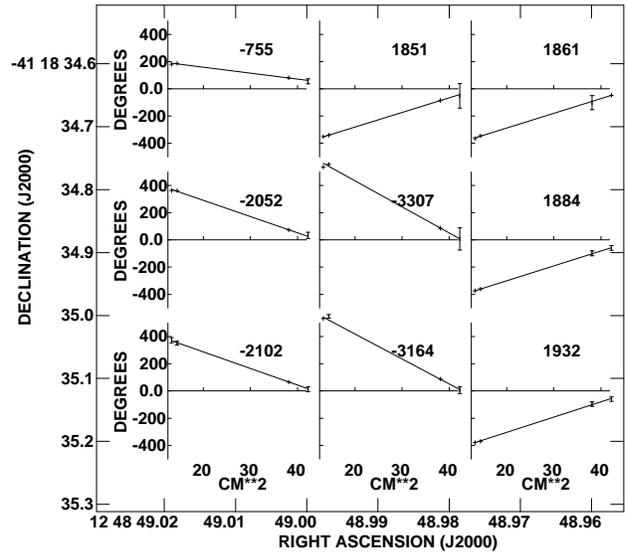,width=0.65\textwidth,angle=-90} }
\caption{Fits to the electric vector position angle as a function
of wavelength squared in the western filament.  The slope of the 
fit provides the RM and is indicated for each pixel.  The transition
from $-$2000 \radm\ to $+$2000 \radm occurs over just 0.6\arcsec.
\label{fig3}}
\end{figure}

\begin{figure*}
\centerline{\psfig{figure=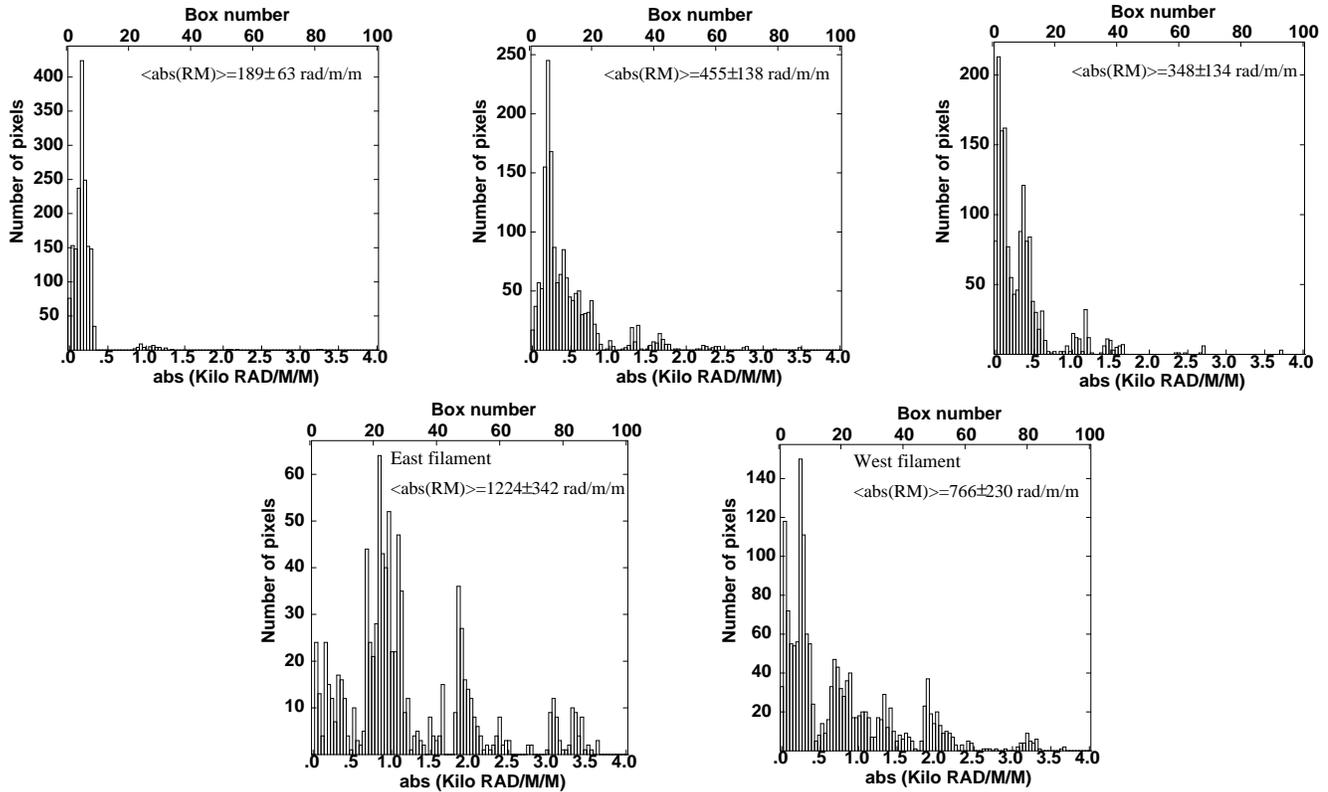,width=1.0\textwidth} }
\caption{Absolute rotation measure distributions for selected areas in 
PKS\,1246{\tt -}410.  The top three regions were selected by taking
a random starting position in the east lobe and then spacing over
at regular intervals of 8.8 arcseconds.
The bottom two regions correspond to the eastern and western 
filaments.  The width of each bin is
40 \radm, which is about the same as the error in the RM determination
for any given point. The number of independent beams is: 11.1, 10.5, and 9.9
respectively for the 3 regions plotted on the top row; and
6.1 and 10.7 for the east and west filament, respectively.
\label{fig4}}
\end{figure*}

\subsection{H$\alpha$/Radio Comparisons}

Depolarization of the radio source coincident with the line-emitting
gas has been seen before (e.g., A1795; Ge \& Owen 1993), and claimed
to be the result of large RM gradients within the synthesised beam.
Even modest magnetic field strengths, combined with the high thermal
densities of the line-emitting gas, can produce large RM gradients.
In Fig.~2 we show overlays of the unsharp-masked H$\alpha$ and the RM
map.  Inspection of Fig.~2 reveals that RM image has values near the
extreme ranges of the image and fewer measurements (presumably due to
depolarization of the radio source) where the H$\alpha$ filaments
cross the lobes.  We find values ranging from $-$2000 \radm\ to
$+$2000 \radm\ on scales of $\sim$ 0.4\arcsec (the maximum resolution
of our observations).  These correspond to gradients of 10$^4$
\radm/arcsec.

A concern about the measurements of the RMs in the filaments is that
they are in regions of relatively low SNR.  To make sure that the high
RMs are not due to low SNR, we show the RM fits in the west filament
in Fig.~3.  The EVPAs are well fit by a $\lambda^2$-law, and have
reasonable reduced $\chi^2$ values of $\sim$1, with a few fits as low
as 0.1 and as high as 10.  To further test the possibility that the
RMs are higher in association with the filaments we compare the
absolute RM distribution in three regularly spaced regions with a
similar area, with those just coincident with the filaments (Fig.~4).
The average $|RM|$ in the east and west filaments is 1224 $\pm$ 342
\radm\ and 766 $\pm$ 230 \radm, compared to averages of below 500
\radm\ for the three typical regions.  We performed a Kolmogorov-Smirnov 
test (KS-test) to see how significant the difference is, and found
that the three typical regions have a $> 99.9\%$ probability of being
drawn from a different population compared to the filament regions.
We also checked for any correlation between low SNR and high RM.  In
general regions of low SNR are equally likely to have any RM, and are
not biased to large absolute values.  We conclude that there is a real
association between the H$\alpha$ emission and the regions of enhanced
RM.
 
The surface brightness of the filaments in H$\alpha + $[NII] is 1-3
$\times$ 10$^{-16}$ erg cm$^{-2}$ s$^{-1}$ arcsec$^{-2}$ (Crawford et
al.\ 2005).  If we take $2 \times 10^{-16}$ erg cm$^{-2}$ s$^{-1}$
arcsec$^{-2}$ for H$\alpha$ in the brightest depolarizing part, 5
arcsec ($\sim$1 kpc) west of the nucleus, and adopt a product $nT = 2
\times 10^6$ cm$^{-3}$ K for the surrounding gas pressure (taking
values from Allen et al. 2006), we find that the depth of the
H$\alpha$ emitting region (T $\sim$ 10$^4$K) is 0.01 pc. This assumes
a uniform covering fraction.  The line-emitting gas is much more
likely to be in the form of a network of small filaments, in which
case the covering fraction is smaller and the depth proportionately
larger, realistically $\sim$1 pc.  If the magnetic fields are
organized on similar scales then the many field reversals within the
synthesized beam (dimensions 252 $\times$ 84 parsec), would scramble
the intrinsic polarization vectors, and lead to depolarization.  Even
if the magnetic fields are for some reason organized on kiloparsec
scales such a network of filaments could not provide the observed high
RM values as they would still produce larger RM gradients than we
observe and would completely depolarize the radio emission due to
variations within the synthesized beam.  Therefore, even though
the H$\alpha$ emission is coincident with regions of enhanced 
RM, it does not have the proper physical characteristics to 
produce coherent RMs on the observed scales.

\begin{figure*}
\centerline{\psfig{figure=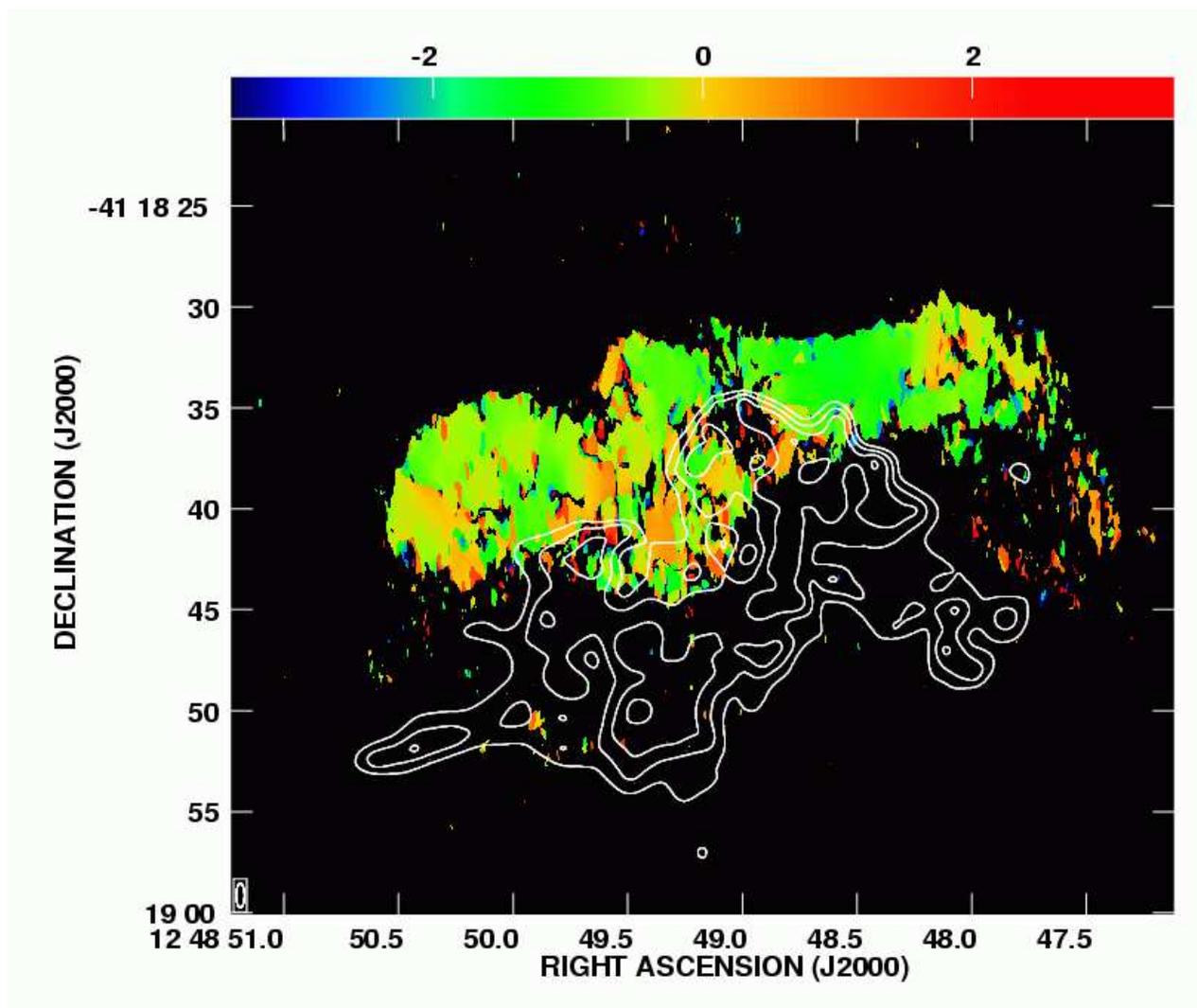,width=0.95\textwidth} }
\caption{Overlay of the Faraday RM on the X-ray temperature map.  
The innermost (highest) contours correspond to temperatures of 
0.5 keV and the lowest contours to 0.9 keV. The temperatures over 
most of the upper part of the radio lobes are around 1.5~keV.
\label{fig5}}
\end{figure*}

\subsection{Soft X-ray/Radio Comparisons}

As pointed out by Crawford et al. (2005) the soft X-ray emitting gas
is coincident with the optical line filaments.  In Fig.~5 we overlay
the X-ray temperature map from Crawford et al. on our Faraday RM
image.  A clear correspondence can be seen between low temperature gas
and the regions of the east and west filaments 
previously noted to have large RM gradients.

A contribution to the RM could come from the dense soft X-ray emitting
gas, of density 0.1 cm$^{-3}$ (Allen et al 2006). This gas has a
covering fraction closer to unity and is much more volume filling than
the line emitting gas outside the radio bubble. Then, if its depth is
about 1 kpc, the magnetic field required to produce the RMs observed
of $\sim$2000
\radm\ is about 25 $\mu$G.  The magnetic pressure is then 2 $\times
10^5$ cm$^{-3}$ K, or about 10 per cent of the thermal pressure.
These values seem reasonable and given the good correspondence with
the soft X-ray emitting gas we believe this is the most likely cause
for the excess RMs and RM gradients observed.  

A simple model where tangled magnetic flux is conserved as gas cools
at constant pressure would predict that $B \propto T^{-2/3}$. This is
in rough agreement with the increase in $B$ from $\sim 11$ to $25\mu$G
expected as the temperature drops from $1.5$ to $\sim 0.6$ keV.  If
this scaling continues to lower temperature the gas at $\sim0.1$ keV
($10^6$ K)
would have a magnetic pressure similar to the thermal pressure. The
optical filaments may well have significant magnetic pressure, likely
needed for maintaining their integrity against shredding by random
motions and conduction. Our observations are however unable to comment
directly on them due to their very small scale (section 6.1).

%

\def\dg{$^{\circ}$}

\begin{table}
\scriptsize
\caption{O{\sc bservational} P{\sc arameters}}
\begin{tabular}{llrrrr}
\hline
\hline
Source & Date & Frequency & Bandwidth$^a$ & Config. & Duration \\
 &  & (GHz) & (MHz) &  & (hours) \\
\hline
PKS\,1246   & Apr1998 & 1465/1565 & 25 & A & 0.24 \\
{\tt -}410         & Jun1998 & 1465/1565 & 25 & BnA & 0.99 \\
         & Apr1998 & 4635/4885 & 50 & A & 0.09 \\
         & Jun1998 & 4635/4885 & 50 & BnA & 1.23 \\
         & Nov1998 & 4635/4885 & 50 & CnB & 0.61 \\
         & Jun1998 & 8115/8485 & 50 & BnA & 0.89 \\
         & Nov1998 & 8115/8485 & 50 & CnB & 0.62 \\
\hline
\end{tabular}
\end{table}

\section{Conclusions}

New VLA observations of the radio galaxy \pks,  providing
high angular and spatial resolution,
afford us with a rare case where we can sample the RM 
structure continuously across the core of a cluster. 
We find distinct
regions, of width $\sim$400 pc, where the RM and RM gradients, are
enhanced, and the fractional polarization is reduced.  This
enhancement in the RM most likely originates in dense, soft X-ray
emitting gas in front of the radio lobe.  The magnetic fields required
to produce the observed enhancement are 25 $\mu$G, similar to the
equipartition estimates for the magnetic fields interior to the lobes,
and a few times larger than the ambient cluster magnetic fields.  If
the magnetic fields continues to increase as the gas cools according
to $B \propto T^{-2/3}$ then the fields may play an important role in
shaping the optical filaments.  To confirm this scenario would require
observations with at least an order of magnitude higher angular
resolution, and a hundredfold increase in sensitivity, such as could
be provided by the proposed Square Kilometer Array.  Observations of
the RM structure in other clusters can be obtained now with the VLA
and some effort, and will soon be made much easier by the advent of
the EVLA.  These radio observations can be compared to sensitive X-ray
images to look for similar correlations between RM enhancements and
soft X-ray emission.

\section{Acknowledgments}

We thank an anonymous referee for constructive suggestions.
GBT acknowledges support for this work from the National Aeronautics
and Space Administration through Chandra Award Numbers GO4-5134X and
GO4-5135X issued by the Chandra X-ray Observatory Center, which is
operated by the Smithsonian Astrophysical Observatory for and on
behalf of the National Aeronautics and Space Administration under
contract NAS8-03060.  ACF thanks The Royal Society for support. 
This research has made use of the NASA/IPAC
Extragalactic Database (NED) which is operated by the Jet Propulsion
Laboratory, Caltech, under contract with NASA.  The National Radio
Astronomy Observatory is a facility of the National Science Foundation
operated under a cooperative agreement by Associated Universities,
Inc.

\end{document}